\documentstyle[preprint,aps]{revtex}
\tightenlines

\newcommand{\ket}[1]{|{#1}\rangle}

\begin{document}

\title{Proposal for the Measurement of Bell-like Correlations from 
Continuous Variables}
\author{T. C. Ralph \cite{ral}, W. J. Munro and R. E. S. Polkinghorne}
\address{Department of Physics, Centre for Laser
Science, University of Queensland, QLD 4072, Australia
}
\maketitle


\begin{abstract}

We show theoretically that Bell-type correlations can be 
observed between continuous variable measurements 
performed on a parametric source. An auxiliary measurement, 
performed on the detection environment, negates the possibility of 
constructing a local realistic description of these correlations. 

\end{abstract}

\vspace{10 mm}

Entanglement is a defining feature of quantum mechanical systems and
leads to correlations between sub-systems of a very non-classical
nature. These in turn lead to fundamentally new interactions and
applications in the field of quantum information \cite{ben95}.
The strange nature of entanglement was first pointed out
by Einstein, Podolsky and Rosen (EPR) \cite{ein35} for the continuous
variables of position and momentum. Though raising philosophical
questions, their formulation did not lead to predictive differences
between quantum mechanics and local realistic theories. An
experimental demonstration of the EPR effect was made by Ou et al
\cite{ou92} by measuring the 2nd order
correlations between the conjugate quadrature
amplitudes of an optical parametric source. The quadrature amplitudes
are the optical analogues of position and momentum and can be measured efficiently 
with homodyne detection. Although the
observed correlations were shown to conflict with semi-classical optical
theory they were not shown to conflict with local realistic theories in 
general.

The deeper mysteries of entanglement were quantified by Bell
\cite{bel65}
in his famous inequalities. These raise testable differences between
quantum mechanics and {\bf all} local realistic theories. Numerous experimental
tests of Bell-type inequalities have been made in optics starting with
Aspect \cite{asp82}. Violations, showing agreement with quantum
mechanics, of over 100 standard deviations \cite{kwi95} and over large
distances \cite{tit98} have now been performed.
In these experiments the correlations between
discrete measurements of particle number are studied. These
correlations are to 4th order in the optical fields.

In this letter we show how Bell-type correlations can be obtained from
measurements of continuous variables. Although defined in terms of
4th order correlations we are able to express our result in terms of
products of only 2nd order correlations. This work is of clear
significance to the new field of continuous variable quantum
information \cite{bra}, but also offers new insights into the
fundamental mechanism of entanglement. Unlike some earlier proposals 
\cite{gra88} our scheme can be applied to macroscopic fields. 
Our proposal differs from previous macroscopic theories
of this kind \cite{gil98} in two ways: (i) The source
upon which we base our quantum mechanical demonstration is a standard
optical parametric amplifier. Previous proposals required the use of
more exotic sources; and (ii) Unlike previous proposals in which the
continuous variables were discretized, we instead use the standard
device of decomposition into two orthogonal polarization bases.
The possibility of constructing a local realistic description of
these correlations, based on the positive Wigner function that describes
the parametric amplifier, can be disallowed by an auxiliary intensity
measurement performed on the detection environment.

Consider the generic correlation experiment shown in Fig.1.
Correlated beams of particles are emitted from a source ($S$)
in opposite directions, $A$ and $B$.
Two distinct paths ($p$ and $m$) are available to the
particles in each beam. These could be different spatial paths
or orthogonal polarizations (or spins), as in
standard realizations. The two paths are combined and then spatially
separated to form a
different pair of orthogonal paths $+$ and $-$. The combiners, $C(\theta)$,
are black boxes such that it is not possible, for a general value of
the mixing parameter $\theta$, to
determine from measurements of $+$ and $-$ whether a particular
particle took path $p$ or $m$. We also allow for a classical phase 
reference (i.e. local oscillators) to be established at the 
measurement sites. Measurements 
are then made on the $+$ and $-$ paths of each beam giving results
$R^{+}(\theta)$
and $R^{-}(\theta)$ respectively. In the standard case these measurements are
simply the presence (1) or absence (0) of a particle in a particular
path in some time interval. More generally they may represent the
count rate of particles in a particular path. We allow for
the most general case in which they may also be constructed from the values of
some continuous properties of the particles (such as position or
momentum) averaged over some time interval. We can form correlation 
functions of the following form
\begin{eqnarray}
R^{ij}(\theta_{A},\theta_{B}) & = &  R_{A}^{i}(\theta_{A})
R_{B}^{j}(\theta_{B}) 
\label{c}
\end{eqnarray}

where  $i,j=+,-$. We then construct the normalized averages
\begin{eqnarray}
P^{ij}(\theta_{A},\theta_{B}) & = &
{{\langle R^{ij}(\theta_{A},\theta_{B})\rangle}\over{
\sum_{k,l=\pm}\langle R^{kl}(\theta_{A},\theta_{B})\rangle}}
\label{p}
\end{eqnarray}
%
%
%
%
It is a well known result \cite{bel71} that provided the $P$'s have the form of
probabilities (bounded between 0 and 1) then in any local realistic
description the correlations will be bounded by the following Bell inequality
\begin{eqnarray}
B=|E(\theta_{A},\theta_{B})+E(\theta_{A}',\theta_{B}')+E(\theta_{A}',\theta_{B})
-
E(\theta_{A},\theta_{B}')| \le 2
\label{ie}
\end{eqnarray}
where
\begin{eqnarray}
E(\theta_{A},\theta_{B})=P^{++}(\theta_{A},\theta_{B})+
P^{--}(\theta_{A},\theta_{B})-P^{+-}(\theta_{A},\theta_{B})-
P^{-+}(\theta_{A},\theta_{B})
\end{eqnarray}
The inequality of Eq. (\ref{ie}) can be applied in the case of the $R$'s
being constructed from continuous measurements provided $
R^{ij} \ge 0$ (and thus $0 \le P^{ij} \le 1$).
It remains to be shown whether there
are any particular continuous measurements which will violate this
inequality for some particular quantum states.

To pursue this goal we first review the standard optical example of a
state which violates this inequality with discrete measurements. Such
a state is the number-polarization entangled state 
$\chi/\sqrt{2}\left( \ket{1_{h},1_{h}} 
+ \ket{1_{v},1_{v}}\right)+\ket{0}$
which is approximately produced by a parametric down converter
operating at low conversion efficiency ($\chi<<1$). Here
$\ket{1_{i},1_{j}}\equiv \ket{1_{i}}_{A}\otimes
\ket{1_{j}}_{B}$ and $1_{h}$ and $1_{v}$ represent single photons in the
horizontal and vertical polarizations respectively. The requirement of
low conversion efficiency is so that higher photon number terms
(which appear as products of higher powers of $\chi$) can be neglected.
In the following it will be more convenient to work in the Heisenberg
picture. In this picture the action of the down converter is to
evolve vacuum state, input annihilation operators $C_{h,v}$ and
$D_{h,v}$ according to
\begin{eqnarray}
\hat A_{h,v}  =  \hat C_{h,v}+\chi \hat D_{h,v}^{\dagger}, \;\;
\hat B_{h,v}  =  \hat D_{h,v}+\chi \hat C_{h,v}^{\dagger}
\label{dc}
\end{eqnarray}
where as before $\chi<<1$ has been assumed.
Our two paths, $p$ and $m$, in this example are the horizontal
($\hat A_{h}$ and $\hat B_{h}$) and vertical ($\hat A_{v}$ and $\hat B_{v}$)
polarization modes.
The mixer $C$ is then some combination of polarizing optics which
decomposes our beams into a different, orthogonal polarization basis
set ($+$ and $-$). This corresponds to the transformation
\begin{eqnarray}
\hat A_{+}(\theta_{A}) = \cos \theta_{A} \hat A_{h}+
\sin \theta_{A} \hat A_{v}, \; \; & &
\hat A_{-}(\theta_{A}) = \cos \theta_{A} \hat A_{v}-
\sin \theta_{A} \hat A_{h}\nonumber\\
\hat B_{+}(\theta_{B}) = \cos \theta_{B} \hat B_{h}+
\sin \theta_{B} \hat B_{v}, \; \; & & 
\hat B_{-}(\theta_{B}) = \cos \theta_{B} \hat B_{v}-
\sin \theta_{B} \hat B_{h}
\end{eqnarray}
Photon counting is then performed on the beams and we define
\begin{eqnarray}
R_{A}^{i}(\theta_{A}) & = & \hat A_{i}^{\dagger}(\theta_{A})
\hat A_{i}(\theta_{A})\nonumber\\
R_{B}^{i}(\theta_{B}) & = & \hat B_{i}^{\dagger}(\theta_{B})
\hat B_{i}(\theta_{B})
\end{eqnarray}
with $i=+,-$.
The definitions then follow as per Eqs. (\ref{c}), and (\ref{p}).
An explicit calculation gives the result
\begin{eqnarray}
E(\theta_{A}, \theta_{B})=\cos 2(\theta_{A}-\theta_{B})
\label{e}
\end{eqnarray}
where terms of order higher than $\chi^{2}$ have been neglected. Choosing the
angles $\theta_{A}=3\pi/8$, $\theta_{A}'=\pi/8$, $\theta_{B}=\pi/4$, $\theta_{B}'=0$ 
we find $B=2 \sqrt{2}$, a clear violation of Eq.(\ref{ie}).

We now consider how we might decompose this result into continuous
variable measurements. The quantum
mechanical properties of correlation functions such as
\begin{eqnarray}
R^{++}=\langle \hat A_{+}^{\dagger}(\theta_{A})
\hat A_{+}(\theta_{A})\hat B_{+}^{\dagger}(\theta_{B})
\hat B_{+}(\theta_{B}\rangle
\label{cc}
\end{eqnarray}
are at the heart of our result.
However, an arbitrary field operator, $\hat F$, can be written as
a sum of the conjugate in-phase, $\hat X_{F;1}=\hat F+\hat F^{\dagger}$, and
out-of-phase, $\hat X_{F;2}=i(\hat F-\hat F^{\dagger})$, quadrature
amplitude operators via $\hat F=1/2(\hat X_{F;1}-i \hat X_{F;2})$.
As noted earlier these operators represent continuous variable
observables which can be measured via homodyne detection with respect 
to local oscillator fields. Thus we can write Eq.(\ref{cc}) in the form
\begin{eqnarray}
R^{++} & = & {{1}\over{16}}\langle (\hat X_{A;1}^{+}(\theta_{A})+i\hat
X_{A;2}^{+}(\theta_{A}))
(\hat X_{A;1}^{+}(\theta_{A})-i\hat
X_{A;2}^{+}(\theta_{A}))\nonumber\\
 & & \times (\hat X_{B;1}^{+}(\theta_{B})+i\hat X_{B;2}^{+}(\theta_{B}))
(\hat X_{B;1}^{+}(\theta_{B})-i\hat X_{B;2}^{+}(\theta_{B}))\rangle
\label{ccx}
\end{eqnarray}
This in turn can be expanded to give the second order correlation
function
\begin{eqnarray}
R^{++} & = & {{1}\over{16}}\left( 2(V_{A;1,B;1}^{+})^{2}+2(V_{A;2,B;2}^{+})^{2}+
2(V_{A;2,B;1}^{+})^{2}
+2(V_{A;1,B;2}^{+})^{2} + \right.\nonumber\\
 & & V_{A;1}^{+}V_{B;1}^{+}+V_{A;2}^{+}V_{B;2}^{+}
 +V_{A;2}^{+}V_{B;1}^{+}+V_{A;1}^{+}V_{B;2}^{+} - \nonumber\\
 & & (1/i)[\hat X_{A;1}^{+}, \hat X_{A;2}^{+}](V_{B;1}^{+}+V_{B;2}^{+})-
 (1/i)[\hat X_{B;1}^{+}, \hat X_{B;2}^{+}](V_{A;1}^{+}+V_{A;2}^{+})-\nonumber\\
 & & \left.[\hat X_{A;1}^{+}, \hat X_{A;2}^{+}]
 [\hat X_{B;1}^{+}, \hat X_{B;2}^{+}])\right)
\label{ccx2}
\end{eqnarray}
where we have assumed Gaussian noise statistics (a valid assumption 
for parametric amplification) and thus expanded 4th
order correlations via $\langle(X_{i}X_{j})^{2}\rangle=
\langle(X_{i})^{2}\rangle \langle(X_{j})^{2}\rangle +2
\langle(X_{i}X_{j})\rangle^{2}$ and defined
$\langle(X_{i}X_{j})\rangle=V_{i,j}$ and
$\langle(X_{i})^{2}\rangle=V_{i}$. The other correlation
functions ($R^{--}, R^{+-}, R^{-+}$) can be formed in a similar way
and hence $E(\theta_{A},\theta_{B})$ constructed.
The terms in the 1st line of Eq.(\ref{ccx2}) are the 4-mode
equivalent (2 spatial $\times$ 2 polarization) of the 2-mode correlations
measured by Ou et al \cite{ou92}. These produce the cosine dependence on
polarizer angle seen in $E(\theta_{A},\theta_{B})$ (Eq.\ref{e}).
The 2nd line terms represent polarization independent noise
which reduces the polarization visibility. The final terms are purely
quantum mechanical, being products with the commutators $[\hat X_{k;1},
\hat X_{k;2}]=2i$ ($k=A,B$). For the down converter these final
terms cancel the 2nd line terms leaving high polarization
visibility, as required to violate the Bell inequality.

Intriguing as this result is Eq.(\ref{ccx2}) does not actually constitute
a continuous variable Bell test as it stands. To do this we must
propose continuous variable measurement protocols, $R_{A}^{+}$ and
$R_{B}^{+}$ from which an $R^{++}=\langle R_{A}^{+}R_{B}^{+}\rangle$ 
can be formed which is
equivalent to Eq.(\ref{ccx2}). 
Consider the following measurement protocol. The observers at $A$ and 
$B$ prearrange synchronized time windows in which they will make 
their measurements. They do not prearrange what measurements they 
will make in a particular window. The observers randomly swap between 
``bright'' measurements of either the in-phase or out-of-phase 
quadratures and ``dark noise'' measurements obtained
by blocking the signal input and allowing {\it no} light to the reach
the homodyne detectors. When the data thus collected is 
brought together the following correlation function can be formed
\begin{eqnarray}
R^{++} & = & \langle ((X_{A;1}^{+})^{2}-(X_{va;1}^{+})^{2}+(X_{A;2}^{+})^{2}-
(X_{va;2}^{+})^{2})\nonumber\\
& & \times \quad
((X_{B;1}^{+})^{2}-(X_{vb;1}^{+})^{2}+(X_{B;2}^{+})^{2}-
(X_{vb:2}^{+})^{2}) \rangle
\label{r}
\end{eqnarray}
where the $X_{vi;j}$ represent the dark noise at the two sites 
($i=a,b$) and on the two quadratures ($j=1,2$). 
Our protocol thus consists of making a series
of homodyne measurements, swapping between the two quadratures,
each of which is then ``zeroed'' by subtracting off the dark noise
of the measurement apparatus. Importantly, for sufficiently long data 
runs, the amount of redundant information will be negligible, i.e. 
{\it all} the data is used in forming the correlation function. 
Similar measurements are made 
on the minus port and the polarization angle is also randomly swapped.
By using the Gaussian properties again we 
find Eq.\ref{r} is equivalent to
\begin{eqnarray}
R^{++} & = &{{1}\over{16}}\left( 2(V_{A;1,B;1}^{+})^{2}+
2(V_{A;2,B;2}^{+})^{2}+2(V_{A;2,B;1}^{+})^{2}
+2(V_{A;1,B;2}^{+})^{2} + \right.\nonumber\\
 & & V_{A;1}^{+}V_{B;1}^{+}+V_{A;2}^{+}V_{B;2}^{+}+V_{A;2}^{+}V_{B;1}^{+}
 +V_{A;1}^{+}V_{B;2}^{+} - \nonumber\\
 & & \left. 2 V_{v}(V_{B;1}^{+}+V_{B;2}^{+})-
 2 V_{v}(V_{A;1}^{+}+V_{A;2}^{+})+4 V_{v}^{2}\right)
 \label{rr}
\end{eqnarray}
Eqs.(\ref{r}) and (\ref{rr}) are the key results of this letter.
They
can be used to construct a test of local realistic theories based on
continuous variable measurements (following the recipe of Eqs.(\ref{p})
and (\ref{ie})). Furthermore it can be shown that
Eqs.(\ref{ccx2}) and (\ref{rr}) are numerically identical.
Thus our continuous variable inequality will be violated by the down
converter. The purely quantum mechanical terms are now those
multiplied by the dark noise ($V_{v}$). For an ideal classical system
this will be zero, leading to low visibility. If dark noise is present
it will also affect the ``bright'' measurements, ensuring no violation
of the Bell inequality. For an ideal quantum mechanical
system the dark noise is produced by vacuum fluctuations (as a
result of non-zero commutation) and will
always be at the quantum noise level ($V_{v}=1$). However for the bright 
measurements the vacuum noise becomes correlated in the quantum mechanical case. 
It is indeed this ability of
entangled states to correlate the
environmental noise which leads to the violation of the predictions of
local realistic theories. It is clear in this formulation that
it is this lack of ``realism'' in the detection process which leads to
the Bell violation. The distribution of the correlations themselves occurs in a
purely local way \cite{deu99}.

It can be shown \cite{ral00} that the correlation function of 
Eq.(\ref{rr}) is formally equivalent to that obtained from the 
discrete measurement $R_{A}^{+}=\hat A^{\dagger}\hat A-\hat 
V^{\dagger}\hat V$ where $\hat V$ is the background vacuum mode. 
Clearly the positivity condition on $R_{A}^{+}$ is preserved provided $\hat V$ 
is truly a vacuum mode. Thus an essential requirement for the validity of 
this test is that the
background measurements are truly ``dark''. This could be ensured by
making a sensitive measurement of the light intensity entering
the homodyne detectors when the signal is blocked. If we make the
reasonable assumption that any stray light will be incoherent, then
a dark photon number satisfying $n_{dark}<<\sqrt{n_{LO}}$ where $n_{LO}$
is the photon number of the local oscillator used to
make the homodyne measurements, can be considered zero. This
auxiliary measurement on the dark input prevents the construction of local
hidden variable theories based on the positive Wigner function which
describes this system. Any hidden variable theory which could successfully
mimic the quadrature correlations would be incompatible with the
observation of zero dark port intensity.

To this point we have demonstrated a
new continuous variable method of measuring non-classical correlations
that have already been shown to exist. We now indicate how
our result can be extended to cover a class of inputs for which local
realistic
violations have not previously been demonstrated. Consider the
arrangement of standard bright squeezed light sources shown in Fig.2.
Four optical parametric amplifiers are seeded by four phase locked,
horizontally polarized
laser beams. The output beams will be squeezed at rf frequencies
high enough such that technical noise on the laser beams can be
neglected but low enough to be within the bandwidth of the amplifiers.
Within this range of frequencies the quantum fluctuations of the
output beams from the squeezers can be described in Fourier space by
the zero-point operators
\begin{eqnarray}
\delta f_{i}=\sqrt{G} \delta g_{i}+\sqrt{G-1} \delta g_{i}^{\dagger}
\end{eqnarray}
where $G$ is the parametric gain of the amplifiers and $\delta
g_{i}$'s are the fluctuations associated with the input beams. These are
assumed to be at the vacuum level for these frequencies. Fourier
space is indicated by the lack of circumflex. These beams are then
combined in the manner shown in Fig.2 to produce 4-mode squeezed
beams (2 spatial $\times$ 2 polarization).
The output beams can be written
\begin{eqnarray}
\delta a_{h,v} & = & \sqrt{G} \delta c_{h,v}+\sqrt{G-1} \delta
d_{h,v}^{\dagger}\nonumber\\
\delta b_{h,v} & = & \sqrt{G} \delta d_{h,v}+\sqrt{G-1} \delta
c_{h,v}^{\dagger}
\label{sq}
\end{eqnarray}
where $\delta c_{h,v}=1/\sqrt{2} (\delta g_{1}+i \delta g_{2})$ and
$\delta d_{h,v}=1/\sqrt{2} (\delta g_{1}-i \delta g_{2})$. For low
levels of parametric gain we can set $G \approx 1$ and
$\sqrt{G-1}=\chi<<1$. Eqs.(\ref{dc}) and (\ref{sq}) are then formally
equivalent although describing physically very different situations.
On one hand Eq.(\ref{dc}) describes the properties of a very low photon number
light beam in the time domain. On the other hand Eq.(\ref{sq})
describes the small fluctuations of a macroscopic light field in the
Fourier domain. These differences don't prevent us constructing
Fourier domain correlation functions completely analogous to those
discussed previously for the down conversion source. Thus the $V$'s
that appear in Eq.(\ref{rr}) are now interpreted as quadrature spectral
variances measured at some rf frequency. In this way it would be
possible to demonstrate correlations between the quadrature amplitude
fluctuations of macroscopic light fields which violate local
realistic theory predictions, a quite remarkable result.

Although the experiment just described is technically challenging it is
certainly within the capabilities of present technology. 
Note that if the parametric gain $G$ becomes too large then higher
order terms will become important and will wash out the non-classical
correlations, i.e. the effect is diluted by too much squeezing. 
This is also typical of the photon number correlations
\cite{wal94}. In Fig.3 we plot the decrease in the maximum value of
$B$ as a function of increasing squeezing. The trade-off with small
levels of squeezing is that the signal to noise becomes very small,
making source stability a critical factor.

We have proposed a method for observing Bell correlations with continuous
variables. We suggest that a Bell inequality violation should be observable
between spatially separated
quadrature fluctuations of a bright source, constructed in a
straightforward manner from squeezed light beams. This research
represents a significant step down the path to realizing all analogues to
discrete quantum information manipulations in continuous variable
systems.

TCR and WJM acknowledges the support of the Australian Research 
Council. 

\begin{figure}
\caption{Schematic of a generic Bell experiment. See text for details}
\end{figure}

\begin{figure}
\caption{Schematic of system for producing polarization/field 
entangled light. Here the beams $\delta f_{1}$ and
$\delta f_{2}$ are combined with a $\pi/2$ phase shift on an asymmetric
unpolarizing beamsplitter to produce the outputs $\delta a_{1}=
1/\sqrt{2} (\delta f_{1}+i
\delta f_{2})$ and $\delta b_{1}=1/\sqrt{2} (\delta f_{1}-i \delta f_{2})$.
The
outputs $f_{3}$ and $f_{4}$ are combined in a similar way forming
outputs $\delta a_{2}$ and $\delta b_{2}$. These latter outputs are
then rotated with half-wave plates into vertical polarization. The
beams $\delta a_{1}$ and $\delta a_{2}$ are then combined on a
polarizing beamsplitter such that they form the two polarization
modes of a single beam. Similarly for $\delta b_{1}$ and $\delta
b_{2}$.}
\end{figure}

\begin{figure}
\caption{Plot of the maximum value of B versus the percentage  
squeezing. A violation is achieved for $B>2$.}
\end{figure}


\end{document}